\documentclass[superscriptaddress,
twocolumn,showpacs,preprintnumbers,amsmath,amssymb]{revtex4}
\usepackage{graphicx}
\usepackage{dcolumn}
\usepackage{bm}
\usepackage{latexsym}
\begin{document}
\title{Dynamic instability of microtubules: effect of catastrophe-suppressing drugs}
\author{Pankaj Kumar Mishra} 
\affiliation{Department of Physics\\
Indian Institute of Technology\\
Kanpur 208016, India.
}
\author{Ambarish Kunwar} 
\affiliation{Department of Physics\\
Indian Institute of Technology\\
Kanpur 208016, India.
}
\author{Sutapa Mukherji} 
\affiliation{Department of Physics\\
Indian Institute of Technology\\
Kanpur 208016, India.
}
\author{Debashish Chowdhury{\footnote{Corresponding author}}} 
\email{debch@iitk.ac.in}
\affiliation{Department of Physics\\ 
Indian Institute of Technology\\
Kanpur 208016, India.
}

\date{\today}%
\begin{abstract} 
Microtubules are stiff filamentary proteins that constitute an 
important component of the cytoskeleton of cells. These are known 
to exhibit a dynamic instability. A steadily growing microtubule 
can suddenly start depolymerizing very rapidly; this phenomenon 
is known as ``catastrophe''. However, often a shrinking microtubule 
is ``rescued'' and starts polymerizing again. Here we develope a 
model for the polymerization-depolymerization dynamics of microtubules 
in the presence of {\it catastrophe-suppressing drugs}. Solving 
the dynamical equations in the steady-state, we derive exact 
analytical expressions for the length distributions of the microtubules 
tipped with drug-bound tubulin subunits as well as those of the 
microtubules, in the growing and shrinking phases, tipped with 
drug-free pure tubulin subunits. We also examine the stability of 
the steady-state solutions. 
\end{abstract}


\maketitle

\section{\label{sec1}Introduction}

Microtubules are filamentary proteins and constitute a major component 
of the cytoskeleton of the eukaryotic cells \cite{alberts,journals}. 
The dynamic cytoskeletal scaffolding not only supports the cell 
architecture and gives rise to changes in the shape of the cell but 
the network of its constituent filamentary proteins also provides 
pathways for intra-cellular transport. In other words, a wide range 
of dynamical processes, which are essential for sustaining life, are 
driven  by the dynamic cytoskeleton. Therefore, a clear theoretical 
understanding of the fundamental physical principles behind the 
polymerization-depolymerization dynamics of the microtubules is 
expected to provide deep insight into the physics of cell shape 
transformations, cell motility, etc. as well as mechanisms of many 
sub-cellular processes like, for example, chromosome segregations 
during mitosis (i.e., cell division). 

{\it Dynamic instability} \cite{mitchison1,mitchison2} is one of the 
unusual non-equilibrium processes that dominate the dynamics of 
microtubule 
polymerization. Each polymerizing microtubule persistently grows for 
a prolonged duration and, then makes a sudden transition to a 
depolymerizing phase; this phenomenon is known as ``catastrophe''. 
However, the subsequent rapid shrinking of a depolymerizing microtubule 
can get arrested when it makes a sudden reverse transition, called 
``rescue'', to a polymerizing phase. The elongation of a microtubule 
takes place by a reversible, non-covalent attachment of a tubulin 
dimer from the tubulin solution. It is now generally believed that 
the dynamic instability of a microtubule is triggered by the loss of 
its {\it guanosine triphoshate} (GTP) cap because of the hydrolysis of 
GTP into {guanosine diphosphate} (GDP). But, the detailed mechanism, 
i.e., how the chemical process of cap loss induces mechanical 
instability, remains far from clear. 

The dynamics of polymerization-depolymerization of microtubules and 
the phenomena of ''catastrophe'' and ''rescue'' 
\cite{erickson,desai,nogales,howard}, have been studied extensively over 
the last decade using simple theoretical models \cite{hill,rubin,dogterom1,dogterom2,fly1,fly2,fly3,jobs,houch,bicout1,bicout2,freed,zimmermann}. 
One of the earliest models of polymerization-depolymerization dynamics 
of microtubules was proposed by Hill \cite{hill} and subsequently 
extended by Rubin and coworkers \cite{rubin,bicout2}. The length of the 
microtubules is {\it discrete} in the Hill model which is formulated in 
terms of two infinite sets of coupled ordinary differential equations. 
Dogterom and Leibler \cite{dogterom1,dogterom2}, however, treated the 
length as a {\it continuous} variable and the dynamical equations 
were reduced to just two coupled partial differential equations.

A large number of different types of antimitotic drug molecules are 
known to bind with free tubulins in solution and/or with tubulins in 
microtubules. The polymerization-depolymerization dynamics of  
microtubules, which play crucial roles in mitosis, is strongly 
influenced by these drugs. The effects of various drug molecules, e.g., 
{\it colchicine}, paclitaxel, vinca alkaloids and {\it taxol}, etc.,  
on the dynamics of microtubules have been investigated experimentally 
for several years \cite{perez, vande1,vande2}, partly because of their  
potential clinical use in combating cancer \cite{peterson,jordan,lobert}. 
These drugs can be broadly classified into two groups. One group consists 
of microtubule-{\it destablizing} agents whereas the members of the other 
group are microtubule-{\it stabilizing} agents.

In this paper we are interested in the effects of microtubule-stabilizing 
agents (i.e., catastrophe-suppressing drugs) on the length distributions 
of the microtubules {\it in the absence of any GTP and GDP in the system}. 
The generic drug molecules of our interest are assumed to bind rapidly 
with free tubulins in solution; when such a tubulin-drug complex binds 
with the growing end of a microtubule, the drug-capped microtubule gets 
stabilized because of the strong suppression of catastrophe phenomenon 
\cite{jordan,lobert}. 

Recall that there are four main rate constants (or, frequencies) that 
characterize the four important processes involved in microtubule 
polymerization/depolymerization dynamics. Two of these are the rate 
constants for the attachment and detachment of tubulin dimers in the 
polymerization and depolymerization phases, respectively while the 
remaining two are the frequencies of catastrophe and rescue. We {\it 
assume} that the generic drug has the following two effects on the 
polymerization dynamics: (i) it reduces the frequency of catastrophe to 
such a large extent that the microtubules capped with drug-bound tubulins 
do not exhibit catastrophe at all, and (ii) it also affects the rate of 
elongation of the microtubules because the rate constant for the attachment 
of a drug-bound tubulin is, in general, different from that of a drug-free 
tubulin. The effects of real catastrophe-suppressing drugs are quite 
complicated and depend also on the dosage of the drug; some even induce 
a ``paused'' phase \cite{jordan}. 

The aim of this paper is to investigate {\it theoretically} the generic 
effects of {\it catastrophe-suppressing} drugs by extending the earlier 
theoretical models, developed by Hill \cite{hill} and Freed \cite{freed}, 
for the dynamics of polymerization of drug-free pure tubulins. We derive 
exact analytical expressions for the steady-state distributions of the 
lengths of the microtubules tipped with the drug-bound tubulin subunits as 
well as those of microtubules tipped with pure (i.e., drug-free) tubulin 
subunits. We carry out linear stability analysis of the steady-state 
distributions and physically interpret the implications of the spectrum 
of the eigenvalues of the stability matrix. 

This paper is organized as follows. In section \ref{sec2} we briefly 
review some of the relevant earlier theoretical models of dynamic 
instability.  In paricular, we summarize the mathematical frameworks 
of the Hill model \cite{hill} and the recent Freed model \cite{freed} 
of dynamic instability of microtubules. In section \ref{sec3} we 
propose an extension of the Hill model so as to capture the effects 
of the catastrophe-suppressing drugs in a simple way. The model is 
made more realistic in section \ref{sec4} by treating the dynamics of 
the concentration of the drug-free tubulins explicitly following a 
Freed-like approach.  The paper ends with a conclusion section \ref{sec5}.

\section{\label{sec2}Brief review of earlier models of dynamic instability}

\subsection{Hill model}

Microtubules consist of $13$ protofilaments, each consisting of 
monomeric units (actually a $\alpha-\beta$ hetero-dimer) each of 
which is approximately $8$ nm long. On the other hand, the  
polymers in the Hill model are one-dimensional. Therefore, some 
authors (see, for example, \cite{govindan}) identify  ``monomeric 
units'' of the one-dimensional Hill model to have a length of 
approximately $8/13 = 0.6$ nm.

Let $P_H^{+}(n,t)$ and $P_H^{-}(n,t)$ be the probabilities of finding a 
microtubule of length $n$, at time $t$, in the growing ($+$) and 
shrinking ($-$) phases, respectively. Moreover, let $P_H(0,t)$ be the 
probability that the MT nucleating site is empty at time $t$. 
Furthermore, we denote the growth rate of the growing microtubules 
and decay rate of the shrinking microtubules by $p^{H}_g$ and $p^{H}_d$, 
respectively, while the frequencies of catastrophes (the transition 
from growing to the shrinking phase) and rescues (the transition 
from the shrinking to growing phase) are denoted by the symbols 
$p^{H}_{+-}$ and $p^{H}_{-+}$, respectively. Interestingly, the four 
parameters $p^{H}_g$, $p^{H}_d$, $p^{H}_{+-}$ and $p^{H}_{-+}$ were 
measured as functions of free tubulin concentration first in 1988 
\cite{walker} by observing single microtubules using video light 
microscopy.  However, the concentration dependence of these parameters, 
if any, does not appear explicitly in the Hill model.

The Master equations for these probabilities are given by \cite{hill} 
\begin{eqnarray}
\frac{dP_H^{+}(n,t)}{dt} = p^{H}_g P_H^{+}(n-1,t) + p^{H}_{-+} P_H^{-}(n,t) - \nonumber \\
(p^{H}_g + p^{H}_{+-}) P_H^{+}(n,t), ~{\rm for}~ n \geq 2, 
\label{eq-hill1}
\end{eqnarray}
\begin{eqnarray}
\frac{dP_H^{-}(n,t)}{dt} = p^{H}_d P_H^{-}(n+1,t) + p^{H}_{+-} P_H^{+}(n,t) - \nonumber \\
(p^{H}_d + p^{H}_{-+}) P_H^{-}(n,t), ~{\rm for}~ n \geq 1, 
\label{eq-hill2}
\end{eqnarray}
\begin{equation}
\frac{dP_H^{+}(1,t)}{dt} = p^{H}_g P_H(0,t) + p^{H}_{-+} P_H^{-}(1,t) - (p^{H}_g + p^{H}_{+-}) P_H^{+}(1,t),  
\label{eq-hill3}
\end{equation} 
and
\begin{equation}
\frac{dP_H(0,t)}{dt} = - p^{H}_g P_H(0,t) + p^{H}_d P_H^{-}(1,t).  
\label{eq-hill4}
\end{equation}
Imposing the normalization 
\begin{equation}
\sum_{n=1}^{\infty} P_H^{+}(n) + \sum_{n=1}^{\infty} P_H^{-}(n) + P_H(0) = 1,
\end{equation}
the steady-state solutions of the equations (\ref{eq-hill1})-(\ref{eq-hill4}) 
are given by \cite{hill}
\begin{equation}
P_H^{+}(n) = x_H^n P_H(0), 
\label{eq-h1}
\end{equation}
and
\begin{equation}
P_H^{-}(n) = x_H^{n-1} y_H P_H(0) 
\label{eq-h2}
\end{equation}
with 
\begin{equation}
P_H(0) = \frac{1-x_H}{1+y_H}, 
\end{equation}
\begin{equation}
x_H = \frac{p^{H}_g(p^{H}_d + p^{H}_{-+})}{p^{H}_d(p^{H}_g + p^{H}_{+-})} 
\label{eq-hillx}
\end{equation}
and 
\begin{equation}
y_H = \frac{p^{H}_g}{p^{H}_d}. 
\label{eq-hilly}
\end{equation}

In order that the distributions $P_H^{\pm} (n)$ are decreasing, rather 
than increasing, functions of $n$ we must have $x_H < 1$ which imposes 
the constraint 
$p^{H}_g (p^{H}_d + p^{H}_{-+}) < p^{H}_d (p^{H}_g + p^{H}_{+-})$, i.e., 
\begin{equation}
p^{H}_g p^{H}_{-+} < p^{H}_d  p^{H}_{+-}
\label{eq-hillstab}
\end{equation}
on the magnitudes of the parameters. Interestingly, in the Dogterom-
Leibler model \cite{dogterom1}, a steady-state characterized by  
exponentially decaying distributions $P^{\pm}$ of the lengths of 
the microtubules is attained only if the condition (\ref{eq-hillstab}) 
is satisfied by the parameters; otherwise, the system never reaches 
any steady-state and mean of the (Gaussian) distribution of the lengths 
of the microtubules continues to increase linearly with time.

\subsection{Freed model}

It has been realized for quite some time \cite{jobs, houch,walker} 
that the rate of growth of the growing microtubules should depend on 
the availability of free tubulin monomers in the solution. However, 
in the Hill model \cite{hill} the kinetic rate equations do not 
involve the  concentration of the tubulin monomers. Recently, 
Freed \cite{freed} has generalized the Hill model to incorporate 
the dependence of the rates on the tubulin monomer concentration. 
Hammele and Zimmermann \cite{zimmermann} carried out independent 
analytical calculations of the same phenomenon by extending the 
Dogterom-Leibler \cite{dogterom1} model. Since our calculations in 
the section \ref{sec4} are based on an extension of Freed's model, 
we summarize here the main points of this approach. 

Suppose, $\rho_0$ is the initial concentration of the tubulin subunits 
and $\rho$ is the corresponding instantaneous concentration at time $t$. 
Similarly, $N_0$ and $N$ are the initial and instantaneous concentrations 
of the free (i.e., without bound tubulin) nucleating sites, respectively. 
The symbols $P_F^{\pm}(n,t)$ in this section denote the {\it concentrations}, 
rather than probabilities, of microtubules in the growing and the 
shrinking phases, respectively. Moreover, binding of a tubulin subunit 
with a free nucleating site takes place at a rate $p^{F}_n$. Using these 
quantities, the kinetic rate equations in the Freed model \cite{freed} 
can now be written as
\begin{eqnarray}
\frac{dP_F^{+}(n,t)}{dt} = p^{F}_g \rho P_F^{+}(n-1,t) + p^{F}_{-+} P_F^{-}(n,t) - \nonumber \\
(p^{F}_g \rho + p^{F}_{+-}) P_F^{+}(n,t), ~{\rm for}~ n \geq 2, 
\end{eqnarray}
\begin{eqnarray}
\frac{dP_F^{-}(n,t)}{dt} = p^{F}_d P_F^{-}(n+1,t) + p^{F}_{+-} P_F^{+}(n,t) - \nonumber \\
(p^{F}_d + p^{F}_{-+}) P_F^{-}(n,t), ~{\rm for}~ n \geq 1, 
\end{eqnarray}
\begin{equation}
\frac{dP_F^{+}(1,t)}{dt} = p^{F}_n \rho N + p^{F}_{-+} P_F^{-}(1,t) - (p^{F}_g \rho + p^{F}_{+-}) P_F^{+}(1,t),  
\end{equation}
and
\begin{equation}
\frac{d\rho}{dt} = - p^{F}_n \rho N - p^{F}_g \rho \sum_{n=1}^{\infty} P_F^{+}(n,t) + p^{F}_d \sum_{n=1}^{\infty} P_F^{-}(n,t). 
\end{equation}
Moreover, tubulin mass conservation imposes the condition 
\begin{equation}
\rho_0 = \rho + Q_F^{+} + Q_F^{-} 
\end{equation}
where 
\begin{equation}
Q_F^{\pm} = \sum_{n=1}^{\infty} n P_F^{\pm}(n,t). 
\end{equation}
Furthermore, 
\begin{equation}
N_0 = N + P_F^{+} + P_F^{-} 
\end{equation}
where 
\begin{equation}
P_F^{\pm} = \sum_{n=1}^{\infty} P_F^{\pm}(n,t). 
\end{equation}

There is one-to-one correspondence between the parameters and dynamical 
variables in the Freed model and those in the Hill model. For example, 
$p_g^F \rho, p_d^F, p_{+-}^F$ and $p_{-+}^F$ correspond to 
$p_g^H, p_d^H, p_{+-}^H$ and $p_{-+}^H$, respectively.

The steady-state solution of this system of kinetic equations is given 
by \cite{chowetal}
\begin{equation}
P_F^{+} = p^{F}_n N_0 \rho (p^{F}_{-+} + p^{F}_d)/D 
\end{equation}
\begin{equation}
P_F^{-} = p^{F}_n N_0 \rho (p^{F}_{+-} + p^{F}_g \rho)/D 
\end{equation}
and 
\begin{equation}
P_F^{-}(1) = \beta' = p^{F}_n N_0 \rho (p^{F}_d p^{F}_{+-} - p^{F}_g \rho p^{F}_{-+})/(p^{F}_d D)
\end{equation}
where 
\begin{equation}
D = p^{F}_n p^{F}_g \rho^2 + (p^{F}_d p^{F}_n + p^{F}_n p_{+-} - p^{F}_g p^{F}_{-+}) \rho + (p^{F}_d p^{F}_{+-} + p^{F}_n \rho p^{F}_{-+}). 
\end{equation}
Finally, using a generating function technique, Freed\cite{freed} 
derived the analytical expressions  
\begin{eqnarray}
P_F^{+}(n)=(a'c')^{(n-1)/2}[(f'+\beta' d') U_{n-1}(\lambda') \nonumber \\ 
- (a'c')^{-1/2}a'f' U_{n-2}(\lambda')] 
\end{eqnarray}
and
\begin{eqnarray}
P_F^{-}(n)=(a'c')^{(n-1)/2}[\beta' U_{n-1}(\lambda') \nonumber \\ 
+ (a'c')^{-1/2}(b'f'-c'\beta') U_{n-2}(\lambda')] 
~~{\rm for}~~ n \geq 2,
\end{eqnarray}
where 
\begin{equation}
a' = (p^{F}_d + p^{F}_{-+})/p^{F}_d,
\label{eq-a'}
\end{equation}
\begin{equation}
b' = -p^{F}_{+-}/p^{F}_d,
\label{eq-b'}
\end{equation}
\begin{equation}
c' = p^{F}_g \rho/(p^{F}_g\rho+p^{F}_{+-}),
\label{eq-c'}
\end{equation} 
\begin{equation}
d' = p^{F}_{-+}/(p^{F}_g\rho + p^{F}_{+-}),
\label{eq-d'}
\end{equation} 
\begin{equation}
f' = p^{F}_n\rho N/(p^{F}_g\rho+p^{F}_{+-}).
\label{eq-f'}
\end{equation}
\begin{equation}
\lambda' = (a' + b'd' + c')/[2(a'c')^{1/2}]
\label{eq-lambda'}
\end{equation}
and
$U_n(\lambda')$ are the Chebyshev polynomial of the second kind 
given by 
\begin{equation} 
U_n(\lambda') = \sin[(n+1)\arccos \lambda']/(1-\lambda'^2)^{1/2}, 
\label{eq-chebyshev}
\end{equation}
together with $U_{-1}(\lambda') = 0$. 

The distributions $P_F^{\pm}(n)$ will be decreasing functions of $n$ 
provided the condition $a'c' < 1$ is satisfied; this condition imposes 
the constraint 
\begin{equation}
p^{F}_g \rho p^{F}_{-+} < p^{F}_d p^{F}_{+-}
\label{eq-freedstab}
\end{equation}
on the magnitudes of the parameters and the value of the drug-free 
tubulin subunits $\rho$ in the steady-state. The condition 
(\ref{eq-freedstab}) becomes identical to (\ref{eq-hillstab}) if 
we identify $p^{F}_g \rho$ with $p^{H}_g$.

While expressing the steady-state distributions $P_{F}^{\pm}(n)$ in 
terms of Chebyshev polynomial, Freed \cite{freed} implicitly assumed 
that $|\lambda'|< 1$. However, as shown in appendix \ref{appA}, $\lambda'$ 
is, in general, larger than unity. We revise the Freed's result in 
section {\ref{sec4} by taking $\lambda' \ge 1$ and, consequently, we 
get a different polynomial instead of the Chebyshev polynomial given 
in (\ref{eq-chebyshev}).

Freed \cite{freed} derived the exact form of the stability matrix. 
we define 
\begin{equation} 
\Delta_{\pm} = \sum_{n=1}^{\infty} \delta P_{\pm}(n) 
\end{equation}
and the column vectors 
\begin{equation}
{\bf V}(t) = \left( \begin{array}{lr} \Delta_+\\
                           \Delta_- \\
                           \delta \rho 
             \end{array} \right)
\end{equation}
\begin{equation}
{\bf N_F} = \left( \begin{array}{lr} 0\\
                           - p_d \\
                               0 
          \end{array} \right)
\end{equation}
The, the equations obtained from the linear stability analysis above 
can be written as 
\begin{equation}
\frac{d{\bf V}(t)}{dt} = {\bf M_F} {\bf V}(t) + {\bf N_F} \delta P^{F}_-(1) 
\end{equation}
where the matrix ${\bf M_F}$ is given by 
\begin{widetext}
\begin{equation}
{\bf M_F} = \left( \begin{array}{llcl} -p_{+-}-p_n[\rho]_{ss}~~ & p_{-+} - p_n[\rho]_{ss}~~ & {\bf M}_{13}\\
                           p_{+-}~~ & - p_{-+}~~ & 0 \\
                           (p_n-p_g)[\rho]_{ss}~~ & p_n[\rho]_{ss}+p_d~~ & {\bf M}_{33} 
          \end{array} \right)
\end{equation}
\end{widetext}
with 
\begin{equation}
{\bf M}_{13} = p_n\{N_0-[P_+]_{ss}-[P_-]_{ss}\}
\end{equation}
and
\begin{equation}
{\bf M}_{33} = (p_n-p_g)[P_+]_{ss}-p_n(N_0-[P_-]_{ss})
\end{equation}

The nature of the stability of the steady-state is determined by the 
eigenvalues of the matrix ${\bf M_F}$. The steady state is stable if all 
the eigenvalues are real and negative. On the other hand, if some 
roots are positive, these would indicate unbounded growth and the 
corresponding steady-state would be unstable. If the characteristic 
equation has a pair of complex conjugate roots then the system will 
either oscillate about the steady state (if the real part is negative) 
or exhibit oscillatory unbounded growth (if the real part is positive). 

We have carried out numerical analysis of this stability matrix and 
obtained the eigenvalues for several sets of values of the model 
parameters; the eigenvalues corresponding to five different values 
of $\rho$, for a fixed set of values of the other model parameters, 
are shown in the table \ref{tab-freed}. Thus, the steady-state 
distribution is stable over the entire range of $\rho$ for the 
chosen set of parameters values.\\

\begin{table}
\begin{tabular}{|c|c|c|c|} \hline
 $\rho$  &$m_1$    & $m_2$  & $m_3$   \\  \hline
  0.4        &-49.984  & -0.282 & -23.719   \\  \hline
  0.8        &-99.999  & -0.295  & -22.507 \\  \hline
  1.6        &-199.999 & -0.299  & -20.456 \\  \hline
  2.0        &-250.000 & -0.299  & -19.566 \\  \hline
  2.5        &-312.500 & -0.300 & -18.577  \\  \hline
\end{tabular}
\caption{The eigenvalues of the linear stability matrix ${\bf M_F}$
in the Freed model. The other common parameters for above table are 
$p_g=125, p_d=900,p_{-+}=0.08, p_{+-}=0.22,  N_0=0.2$ (in respective units).
}
\label{tab-freed}
\end{table}

\section{\label{sec3}Hill-like model with drugs}

Let $p^{h}_c$ be the rate of growth of a microtubule by addition of a 
drug-bound tubulin.  Since addition  of catastrophe-suppressing 
drugs to the system strongly reduce the catastrophe frequency, 
{\it we assume that the drug is such that it arrests catastrophe.
In other words, a microtubule tipped with a drug-bound tubulin can grow
but cannot shrink}. We shall use the symbol $\Pi_{h}(n,t)$ to denote 
the probability of a microtubule, tipped with a drug-bound tubulin, 
that has length $n$ at time $t$. As a consequence of our assumption, 
we do not need to consider the two quantities $\Pi_h^{+}(n,t)$ and 
$\Pi_h^{-}(n,t)$ separately; $\Pi_h^{-}(n,t) = 0$ for all $n$ at all $t$. 
However, even in the presence of such drugs in the system, catastrophes 
can take place in microtubules tipped with drug-free tubulins. The 
distributions of the microtubules tipped with drug-free tubulin 
subunits in the growing and shrinking phases are denoted by 
$P_h^{+}(n,t)$ $P_h^{-}(n,t)$, respectively.

\begin{figure}[h]
\begin{center}
\includegraphics[width=0.78\columnwidth]{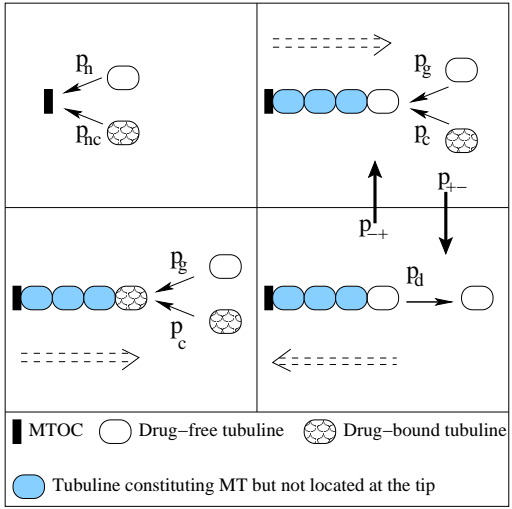}
\end{center}
\caption{A schematic description of our Hill-like model in the presence 
of catastrophe-suppressing drugs. The superscript $h$ has been dropped 
from the symbols denoting the model parameters. 
}
\label{fig-model}
\end{figure}

Thus, the equations (\ref{eq-hill1})-(\ref{eq-hill4}) are generalized 
to the forms 
\begin{eqnarray}
\frac{dP_h^{+}(n,t)}{dt} &=& p^{h}_g [P_h^{+}(n-1,t) + \Pi_h(n-1,t)] + p^{h}_{-+} P_h^{-}(n,t) \nonumber \\
&-& (p^{h}_g + p^{h}_c + p^{h}_{+-}) P_h^{+}(n,t), ~{\rm for} ~n \geq 2, \nonumber \\
\label{eq-htype1}
\end{eqnarray}
\begin{eqnarray}
\frac{dP_h^{-}(n,t)}{dt} = p^{h}_d P_h^{-}(n+1,t) + p^{h}_{+-} P_h^{+}(n,t)  \nonumber \\ 
- (p^{h}_d + p^{h}_{-+}) P_h^{-}(n,t), ~{\rm for}~ n \geq 1, 
\label{eq-htype2}
\end{eqnarray}
\begin{equation}
\frac{dP_h^{+}(1,t)}{dt} = p^{h}_g P_h(0,t) + p^{h}_{-+} P_h^{-}(1,t) - (p^{h}_g + p^{h}_c + p^{h}_{+-}) P_h^{+}(1,t),  
\label{eq-htype3}
\end{equation}
\begin{equation}
\frac{dP_h(0,t)}{dt} = - (p^{h}_g + p^{h}_c) P_h(0,t) + p^{h}_d P_h^{-}(1,t),  
\label{eq-htype4}
\end{equation}
\begin{eqnarray}
\frac{d\Pi_h(n,t)}{dt} &=& p^{h}_c [\Pi_h(n-1,t) + P_h^{+}(n-1,t)] \nonumber \\ 
&-& (p^{h}_g + p^{h}_c) \Pi_h(n,t),  ~{\rm for} ~n \geq 2, \nonumber \\  
\label{eq-htype5}
\end{eqnarray}
\begin{equation}
\frac{d\Pi_h(1,t)}{dt} = p^{h}_c P_h(0,t) - (p^{h}_g + p^{h}_c) \Pi_h(1,t).   
\label{eq-htype6}
\end{equation}
In order to distinguish between the Hill model and our Hill-like 
model with catastrophe-suppressing drugs, we replace the subscripts 
(and superscripts) $H$ of the former by $h$ in the latter.

The steady-state solutions of these kinetic equations (see appendix 
\ref{appB} for details) are 
\begin{equation}
P_h^{+}(n) = x_h (x_h+z_h)^{n-1} P_h(0), 
\label{eq-hsol1}
\end{equation}
\begin{equation}
P_h^{-}(n) = y_h (x_h+z_h)^{n-1} P_h(0), 
\label{eq-hsol2}
\end{equation}
\begin{equation}
\Pi_h(n) = z_h (x_h+z_h)^{n-1} P_h(0), 
\label{eq-hsol3}
\end{equation}
where 
\begin{equation}
P_h(0) = \frac{1-x_h-z_h}{1+y_h} 
\label{eq-hsol4}
\end{equation}
with 
\begin{equation}
x_h = \frac{p^{h}_g(p^{h}_d + p^{h}_{-+})+p^{h}_c p^{h}_{-+}}{p^{h}_d(p^{h}_g + p^{h}_{+-})+p^{h}_c p^{h}_d} 
\label{eq-htypex}
\end{equation} 
and
\begin{equation}
y_h = \frac{p^{h}_g+p^{h}_c}{p^{h}_d} 
\label{eq-htypey}
\end{equation}
\begin{equation}
z_h = \frac{p^{h}_c}{p^{h}_c + p^{h}_g} 
\label{eq-htypez}
\end{equation}

\begin{figure}[h]
\begin{center}
\includegraphics[width=0.75\columnwidth]{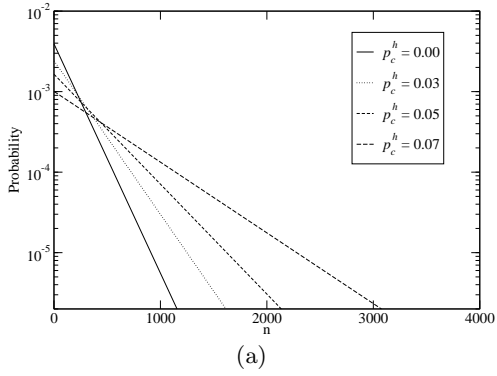} 
\vspace{.75cm}
\centerline{(a)}
\includegraphics[width=0.75\columnwidth]{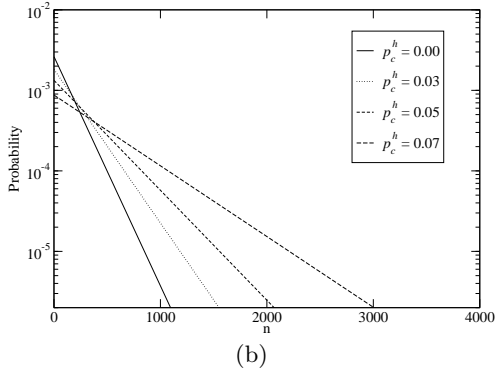}
\vspace{.75cm}
\centerline{(b)}
\includegraphics[width=0.75\columnwidth]{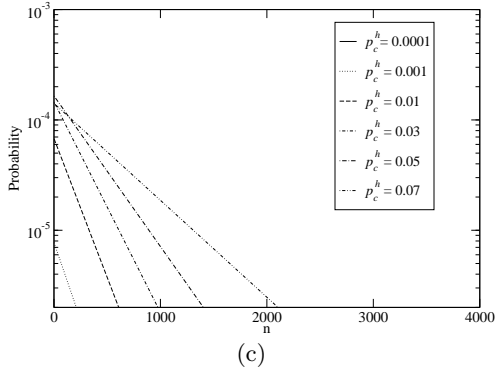}
\vspace{.75cm}
\centerline{(c)}
\end{center}
\caption{The steady-state distributions (a) $P_{h}^{+}(n)$, 
(b) $P_{h}^{-}(n)$ and (c) $\Pi_{h}(n)$ of the size $n$ of 
microtubules in our Hill-like model in the presence of 
catastrophe-suppressing drugs are plotted for several values 
of $p^{h}_c$; the common parameters are 
$p^{h}_g=0.5, p^{h}_d=0.75, p^{h}_{+-}=0.01$ and $p^{h}_{-+}=0.01$, 
each  {\it per unit time}.  
}
\label{fig-hilllike}
\end{figure}

The distributions (\ref{eq-htypex}-\ref{eq-htypez}) are decreasing 
functions of $n$ provided $(x_h + z_h) < 1$, i.e, 
\begin{equation}
(p^{h}_g + p^{h}_c)^2 ~p^{h}_{-+} < p^{h}_g ~p^{h}_d ~p^{h}_{+-}
\label{eq-hlikestab}
\end{equation}
this condition (\ref{eq-hlikestab}) reduces to the condition 
(\ref{eq-hillstab}) in the limit $p^{h}_c \rightarrow 0$.
Note that in the limit $p^{h}_c \rightarrow 0$, $z_h \rightarrow 0$ while 
the expressions (\ref{eq-htypex}) and (\ref{eq-htypey}) for $x_h$ and 
$y_h$ reduce to $x_H$ and $y_H$ given by the expressions (\ref{eq-hillx}) 
and (\ref{eq-hilly}), respectively; hence, in this limit, 
$\Pi_h(n) \rightarrow 0$ and the expressions (\ref{eq-hsol1}) and 
(\ref{eq-hsol2}) for $P_h^{\pm}(n)$ reduce to the corresponding 
expressions (\ref{eq-h1}) and (\ref{eq-h2}) for $P_H^{\pm}(n)$. 

The parameters $p^{h}_g$, $p^{h}_d$, $p^{h}_{+-}, p^{h}_{-+}$, 
etc. have dimensions $[time^{-1}]$. For a typical set of values of these 
parameters, we have plotted the distributions $P_h^{+}(n)$, $P_h^{-}(n)$ and 
$\Pi_h(n)$ in fig.\ref{fig-hilllike}(a),(b) and (c), respectively, each 
for several different numerical values of $p^{h}_c$. The straight 
lines on the semi-log plots is a reflection of the exponential decay 
of the distribution with increasing length of the microtubules. Moreover, 
the longer tails of these distributions at at higher values of $p^{h}_c$ 
demonstrates that higher $p^{h}_c$ causes stronger suppression of the 
{\it catastrophe} phenomenon. 

\subsection{Stability of the steady-state} 

Let us define the small deviations
\begin{equation}
\delta P_{h}^{\pm}(n) = P_{h}^{\pm}(n,t) - [P_{h}^{\pm}]_{ss}
\end{equation}
\begin{equation}
\delta \Pi_{h} (n) = \Pi_{h} (n,t) - [\Pi_{h}]_{ss}
\end{equation}
from the corresponding steady-states where the steady-state values
$[P_{\pm}]_{ss}$ and $[\Pi]_{ss}$ are given by the equations
(\ref{eq-hsol1}), (\ref{eq-hsol2}) and (\ref{eq-hsol3}), respectively.
Linear stability analysis of the steady-state distributions leads to 
the equations  
\begin{eqnarray} 
\frac{d\delta P_{h}^{+}(n,t)}{dt} = p_{g}^{h} [\delta P_{h}^{+}(n-1,t) + \delta \Pi_{h}(n-1,t)] \nonumber \\
+(p_{g}^{h} + p_{c}^{h}+p_{+-}^{h}) \delta P_{h}^{+}(n,t)~~ ~~{\rm for}~~ n \geq 2
\end{eqnarray}
\begin{eqnarray} 
\frac{d \delta P_{h}^{-}(n,t)}{dt} = p_{d}^{h} \delta P_{h}^{-}(n+1,t) + p_{+-}^{h} \delta P_{h}^{+}(n,t) \nonumber \\
- (p_{d}^{h} + p_{-+}^{h}) \delta P_{h}^{-}(n,t) \nonumber \\
{\rm for}~~ n \geq 1
\end{eqnarray}
\begin{eqnarray} 
\frac{d \delta \Pi_{h}(n,t)}{dt} = p_{c}^{h} \{\delta \Pi_{h} (n-1,t) + \delta P_{h}^{+}(n-1,t)\} \nonumber \\
- (p_{g}^{h}+p_{c}^{h}) \delta \Pi_{h}(n,t) {\rm for}~~ n \geq 2
\end{eqnarray}
\begin{eqnarray} 
\frac{d \delta P_{h}^{+}(1,t)}{dt} = p_{g}^{h} \delta P(0,t)+ p_{-+}^{h} \delta P_{h}^{-}(1,t) \nonumber \\ 
- (p_{g}^{h}+p_{c}^{h} + p_{+-}^{h}) \delta P_{h}^{+}(1,t) 
\end{eqnarray}
\begin{eqnarray} 
\frac{d \delta \Pi_{h}(1,t)}{dt} = p_{c}^{h} \delta P(0,t) - (p_{g}^{h}+p_{c}^{h}) \delta \Pi_{h}(1,t)\} \nonumber \\ 
\end{eqnarray}

Using the normalization condition we get
\begin{equation}
\delta P(0,t)= -\sum_{n=1}^{\infty} \delta P_{h}^{+}(n,t) 
-\sum_{n=1}^{\infty} \delta P_{h}^{-}(n,t)
-\sum_{n=1}^{\infty} \delta  \Pi_{h}(n,t) 
\end{equation}
Next, we define 
\begin{eqnarray} 
\Delta_{\pm} = \sum_{n=1}^{\infty} \delta P_{h}^{\pm}(n) \\
\Delta_0 = \sum_{n=1}^{\infty} \delta \Pi_{h}(n) 
\end{eqnarray}
and the column vectors 
\begin{equation}
{\bf W}(t) = \left( \begin{array}{lr} \Delta_+\\
                           \Delta_- \\
                           \Delta_0 \\
             \end{array} \right)
\end{equation}
\begin{equation}
{\bf N_h} = \left( \begin{array}{lr} 0\\
                           - p_{d}^{h} \\
                               0 \\
          \end{array} \right)
\end{equation}
The equations obtained from the linear stability analysis above 
can be written as 
\begin{equation}
\frac{d{\bf W}(t)}{dt} = {\bf M_h} {\bf W}(t) + {\bf N_h} \delta P_{h}^{-}(1,t) 
\end{equation}
where the matrix ${\bf M_h}$ is given by 
\begin{widetext}
\begin{equation}
{\bf M_h} = \left( \begin{array}{llcl} - (p_{c}^{h} + p_{+-}^{h}-p_{g}^{h})~~ & p_{-+}^{h} - p_{g}^{h}~~ & 0\\
                p_{+-}^{h}~~ & - p_{-+}^{h}~~ & 0 \\
                0~~ & - p_{c}~~ & -(p_{g}^{h}+p_{c}^{h}) 
          \end{array} \right)
\end{equation}
\end{widetext}

\vspace{1cm}

\begin{table}
\begin{tabular}{|c|c|c|c|c|} \hline
 $p_{c}^{h}$ &      $m_1$     &      $m_2$      &   $m_3$  \\ \hline
   0   &-100.000        & -100.000        & -0.300  \\ \hline
  10   &-100.020        & -100.000        & -0.280  \\ \hline
  20   &-120.037        & -120.000        & -0.263  \\ \hline
  30   &-130.051        & -130.000        & -0.249  \\ \hline
  50   &-150.073        & -150.000        & -0.227  \\ \hline
 100   &-200.110        & -200.000        & -0.190  \\ \hline
 150   &-250.132        & -250.000        & -0.168  \\ \hline
\end{tabular}
\caption{The eigenvalues of the linear stability matrix ${\bf M_h}$ 
in our Hill-like model. The common parameters are
$p_{g}^{h}=100, p_{d}^{h}=900, p_{-+}^{h}=0.08, p_{+-}^{h}=0.22$, 
(each per {\it unit time})}. 
\label{tab-1}
\end{table}
Note that all the eigenvalues remain negative up to a reasonably high 
value of $p_c$ corresponding to the parameter set chosen for the table 
\ref{tab-1}. This result indicates that the steady state distributions 
of the lengths of the microtubules, which we have derived in this 
section, remain stable up to a moderately high dosage of the catastrophe 
suppressing drug.

\section{\label{sec4}Effects of drugs on MT: A hybrid Hill-Freed-like model}

In this section we extend the Hill-like model developed in section 
\ref{sec3} by taking into account the dependence of the rate of 
growth of the microtubules on the concentration of the drug-free 
tubulin subunits in the solution; for this purpose we follow the 
corresponding approach developed by Freed (and reviewed in section 
\ref{sec2}) for microtubule dynamics in the absence of drugs. However, 
the effects of the drug-bound tubulin subunits are taken into account 
in the same way as was done in the Hill-like model presented in the 
section \ref{sec3}. Therefore, the model presented in this section 
is a {\it hybrid of Hill-like and Freed-like approaches}; the drug-free 
tubulins are treated following Freed while the drug-bound tubulins 
are treated following Hill. 

The binding of a drug-bound tubulin subunit with a free nucleating 
site takes place with probability $p_{nc}$ per unit time. Following 
Freed, the {\it concentration} of microtubules of length $n$ which are 
tipped with drug-bound tubulin subunits is denoted by the symbol 
$\Pi(n,t)$ while that of the microtubules tipped with drug-free 
tubulin subunits and in the growing (shrinking) phase is denoted by 
$P_{+}(n,t)$ ($P_{-}(n,t)$). Moreover, all the parameters with 
identical susbcripts in this model and in the Freed model have the 
same physical significance. The equations of our interest are 
\begin{eqnarray}
\frac{dP_+(n,t)}{dt} = p_g \rho [P_+(n-1,t) + \Pi(n-1,t)] + p_{-+} P_-(n,t) \nonumber \\
- (p_g \rho + p_c + p_{+-}) P_+(n,t) ~~{\rm for}~~ n \geq 2,
\label{eq-hyb1}
\end{eqnarray}
\begin{eqnarray}
\frac{dP_-(n,t)}{dt} = p_d P_-(n+1,t) + p_{+-} P_+(n,t) \nonumber \\
- (p_d + p_{-+}) P_-(n,t)  ~~{\rm for}~~ n \geq 1,
\label{eq-hyb2}
\end{eqnarray}
\begin{eqnarray}
\frac{d\Pi(n,t)}{dt} = p_c [\Pi(n-1,t) + P_+(n-1,t)] \nonumber \\
- (p_g \rho + p_c) \Pi(n,t) ~~{\rm for}~~ n \geq 2,
\label{eq-hyb3}
\end{eqnarray}
\begin{eqnarray}
\frac{dP_+(1,t)}{dt} = p_n \rho N + p_{-+} P_-(1,t) \nonumber \\
- (p_g \rho + p_c + p_{+-}) P_+(1,t),
\label{eq-hyb4}
\end{eqnarray}
\begin{equation}
\frac{d\Pi(1,t)}{dt} = p_{nc} N  - (p_g \rho + p_c) \Pi(1,t),
\label{eq-hyb5}
\end{equation}
and 
\begin{equation}
\frac{d\rho}{dt} = - p_n \rho N - p_g \rho \sum_{n=1}^{\infty} 
[P_+(n,t) + \Pi(n,t)] + p_d \sum_{n=1}^{\infty} P_-(n,t).
\label{eq-hyb6}
\end{equation}

The steady state equations are 
\begin{eqnarray}
p_n \rho (N_0-P_+-P_- -\Pi)=-p_g \rho (P_+ + \Pi) + p_d P_-,
\label{concsteady}
\end{eqnarray}
\begin{eqnarray}
P_+(n)= c P_+(n-1)+d P_-(n)+ c \Pi(n-1),
\label{pplsteady}
\end{eqnarray}
\begin{eqnarray}
P_-(n)=a P_-(n-1)+bP_+(n-1), 
\label{pminsteady}
\end{eqnarray}
\begin{eqnarray}
\Pi(n)=e \Pi(n-1)+e P_+(n-1),
\label{pisteady}
\end{eqnarray}
\begin{eqnarray}
P_+(1)=d P_-(1)+f, 
\label{ppl1steady}
\end{eqnarray}
\begin{eqnarray}
\Pi(1)=g 
\label{pi1steady},
\end{eqnarray}
where 
\begin{equation}
a=(p_d+p_{-+})/p_d,  
\label{eq-a}
\end{equation}
\begin{equation}
b=-p_{+-}/p_d,  
\label{eq-b}
\end{equation}
\begin{equation}
c=p_g \rho/(p_g \rho +p_c+p_{+-}), 
\label{eq-c}
\end{equation}
\begin{equation}
d=p_{-+}/(p_g \rho+p_c+p_{+-}),  
\label{eq-d}
\end{equation}
\begin{equation}
e=p_c/(p_g \rho+p_c),
\label{eq-e}
\end{equation}
\begin{equation}
f=p_n \rho N/(p_g \rho+p_c+p_{+-}), 
\label{eq-f}
\end{equation} 
and
\begin{equation}
g=p_{nc} N/(p_g \rho+p_c).
\label{eq-g}
\end{equation} 
Note that in the limit of vanishing concentration of drug-bound 
tubulin, i.e., $p_c \rightarrow 0$, the expressions (\ref{eq-a})-
(\ref{eq-d}) for $a, b, c,$ and $d$ reduce to the expressions 
(\ref{eq-a'})-(\ref{eq-d'}) for $a', b', c'$ and $d'$, respectively. 
Moreover, in the limit $p_c \rightarrow 0$ and $p_{nc} \rightarrow 0$, 
equations (\ref{eq-e}) and (\ref{eq-g}) imply $e \rightarrow 0$ 
and $g \rightarrow 0$ while the expression (\ref{eq-f}) reduces to 
the corresponding expression (\ref{eq-f'}) of the original Freed model.

Defining 
\begin{eqnarray}
P_+=\sum_{n=1}^\infty P_+(n), \nonumber \\
P_-=\sum_{n=1}^\infty P_-(n),\nonumber \\
\Pi=\sum_{n=1}^\infty \Pi(n), 
\label{eq-sumps}
\end{eqnarray}
we obtain the following three equations 
from (\ref{pminsteady}), (\ref{pplsteady}) and (\ref{pisteady}),
\begin{eqnarray}
(1-a) P_- -b P_+= P_-(1),\label{pminsum}\\
(1-c)P_+ -c \Pi -d P_- =f,\label{pplsum}\\
(1-e) \Pi-e P_+=\Pi(1),\label{pisum}
\end{eqnarray}        
We solve for 
$P_+$, $P_-$ and $\Pi$ using 
(\ref{pplsum}), (\ref{pisum}) and (\ref{concsteady}). 
Substituting these solutions in (\ref{pminsum}),
we obtain $P_-(1)$ at the steady state. Finally, steady state expressions 
for $P_+(1)$ and $\Pi(1)$ are obtained from   
(\ref{ppl1steady}) and (\ref{pi1steady}).
 
The solutions of linear equations (\ref{pplsum}), (\ref{pisum}) and 
(\ref{concsteady}) can be obtained in a straightforward way. The 
solutions are 
\begin{eqnarray}
P_+=\frac{\rho N_0 p_g p_2(\rho p_n+p_{nc})}{{\cal D}}, 
\label{ppl}
\end{eqnarray}
\begin{eqnarray}
P_-=\frac{\rho N_0 p_g (\rho p_n+p_{nc})(p_1+\rho p_g)}{{\cal D}}, 
\label{pminsoln}
\end{eqnarray}
\begin{eqnarray}
\Pi=\frac{N_0[\rho p_c p_np_2-\rho p_g p_{-+} p_{nc}+p_d p_{nc}p_1]}{{\cal D}}, 
\label{pisoln}
\end{eqnarray}
\begin{eqnarray}
P_-(1)=\frac{\rho N_0 p_g(\rho p_n+p_{nc})(p_{+-} p_d-p_c p_{-+}-
\rho p_{-+} p_g)}{( p_d {\cal D})},\nonumber \\
\end{eqnarray}
where 
\begin{eqnarray}
{\cal D}=\rho^3 p_g^2 p_n+p_dp_{nc}p_1 +\rho^2 p_g^2 (p_{nc}-p_{-+}) \nonumber \\ 
+ \rho^2 p_g p_n(p_1+p_2)+\rho p_c p_2 p_n +\rho p_g p_c(p_{nc}-p_{-+}) \nonumber \\ 
+\rho p_g p_{nc} p_{+-}+\rho p_g p_d (p_{nc}+p_{+-}),
\end{eqnarray}
with 
\begin{equation}
p_1 = p_c + p_{+-}
\end{equation}
and
\begin{equation}
p_2 = p_d + p_{-+}.
\end{equation}

In the following the generating function method is used to obtain the 
$P_+(n)$, $P_-(n)$ and $\Pi(n)$ for arbitrary $n$. We proceed by 
multiplying  both sides of  equations (\ref{pplsteady}), 
(\ref{pminsteady}) and (\ref{pisteady}) by $x^n$ and then suming over 
$n$ from $n=1$ to $\infty$.  Defining 
\begin{equation}
P_+(x)=\sum_{n=1}^\infty P_+(n) x^n, 
\end{equation}
\begin{equation}
P_-(x)=\sum_{n=1}^\infty P_-(n) x^n 
\end{equation}
and 
\begin{equation}
\Pi(x)=\sum_{n=1}^\infty \Pi(n) x^n, 
\end{equation}
we have the following equations for $P_+(x)$, $P_-(x)$ and $\Pi(x)$:
\begin{eqnarray}
(1-c x)P_+(x)-d P_-(x)-c x \Pi(x)=x f 
\label{pplgen}
\end{eqnarray}
\begin{eqnarray}
b x P_+(x)+(a x-1) P_-(x)=-x \beta 
\label{genpmin}
\end{eqnarray}
\begin{eqnarray}
e x P_+(x)+(e x-1) \Pi(x)=-x g 
\label{genpi}
\end{eqnarray}
where $\beta = P_-(1)$.
Solving this set of linear equations for $P_+(x)$, $P_-(x)$ and $\Pi(x)$,
we find 
\begin{eqnarray}
&&P_+(x) = \nonumber \\
&&\frac{(aef-acg) x^3-(\beta de-cg+ef+af) x^2 +(\beta d+f)x}{\Delta},\nonumber \\
\label{pplx}
\end{eqnarray}
\begin{eqnarray}
&&P_-(x)=\nonumber \\
&&\frac{(bcg-bef)x^3+(bf-\beta c-\beta e)x^2+\beta x}{\Delta},\nonumber \\
\label{pmix}
\end{eqnarray}
\begin{eqnarray}
&&\Pi(x)=\nonumber \\
&&\frac{(acg-aef)x^3+(\beta de-cg+ef-ag-bdg)x^2+gx}{\Delta}.\nonumber \\
\label{pix}
\end{eqnarray}
where 
\begin{equation}
\Delta = (ac+ae+bde)x^2-(a+c+e+bd)x+1.
\label{eq-delta}
\end{equation}

\begin{eqnarray}
&&P_+(x) = \nonumber \\
&&\frac{(aef-acg) x^3-(\beta de-cg+ef+af) x^2 +(\beta d+f)x}
{y^2-2\lambda y+1},\nonumber \\
\label{pplx}
\end{eqnarray}
\begin{eqnarray}
&&P_-(x)=\nonumber \\
&&\frac{(bcg-bef)x^3+(bf-\beta c-\beta e)x^2+\beta x}
{y^2-2\lambda y+1},\nonumber \\
\label{pmix}
\end{eqnarray}
\begin{eqnarray}
&&\Pi(x)=\nonumber \\
&&\frac{(acg-aef)x^3+(\beta de-cg+ef-ag-bdg)x^2+gx}
{y^2-2\lambda y+1}.\nonumber \\
\label{pix}
\end{eqnarray}
where
\begin{equation}
\lambda=(a+e+c+bd)/[2 (ac+ae+bde)^{1/2}].
\end{equation}
and 
\begin{equation}
y=(a+e+c+bd)^{1/2}x.
\end{equation}

Solutions for $P_{\pm}(n)$ and $\Pi(n)$ can be obtained by expanding 
term 
\begin{equation}
\frac{1}{y^2-2\lambda y+1}
\end{equation}
in the above equations using Taylor series expansion for $\lambda \ge 1$ 
and $y<1$ and then equating the coefficients of $y^n$ on both sides.
Thus, the expressions for $P_{\pm}(n)$ and $\Pi(n)$ are 
as follows:
\begin{eqnarray}
P_+(n)=\alpha^{(n-1)/2}[\alpha_3 U_{n-1}(\lambda)-\alpha_2 U_{n-2}(\lambda)+\alpha_1 U_{n-3}(\lambda)] \nonumber \\
~~{\rm for}~~ n \geq 2, \nonumber \\
\label{eq-hybppn}
\end{eqnarray}
\begin{equation}
P_+(1)=\alpha_3 U_0(\lambda) 
\label{eq-hybpp1}
\end{equation}
where 
\begin{eqnarray}
\alpha=(ac+ae+bde),\\
\alpha_1=(aef-acg)/\alpha,\\
\alpha_2=(\beta de-cg+ef+af)/\alpha^{1/2},\\
\alpha_3=(\beta d+f),
\end{eqnarray}
and 
\begin{equation}
U_{n}(\lambda)=\sum_{m=0}^{[n/2]}(-1)^{m}\frac{(n-m)!}{m!(n-2m)!}(2\lambda)^{n-2m},
\label{new}
\end{equation}
with $U_{-1}(\lambda) = 0$, where the symbol $[n/2]$ represents the 
largest integer smaller than or equal to $n/2$.
Similarly,
\begin{eqnarray}
P_-(n)=\alpha^{(n-1)/2}[\beta U_{n-1}(\lambda)+ \beta_2 U_{n-2}(\lambda)+\beta_1 U_{n-3}(\lambda)] \nonumber \\
~~{\rm for}~~ n \geq 2, \nonumber \\
\label{eq-hybpmn}
\end{eqnarray}
\begin{equation}
P_-(1)=\beta U_0(\lambda) 
\label{eq-hybpm1}
\end{equation}
where 
\begin{eqnarray}
\beta_1=(bcg-bef))/\alpha\\
\beta_2=(bf-\beta c-\beta e)/\alpha^{1/2}.
\end{eqnarray} 
Finally, 
\begin{equation}
\Pi(n)=\alpha^{(n-1)/2}[g U_{n-1}(\lambda)+\gamma_2 U_{n-2}(\lambda)+
\gamma_1 U_{n-3}(\lambda)],
\label{eq-hybpin}
\end{equation}
\begin{equation}
\Pi(1)=g U_0(\lambda)
\label{eq-hybpi1}
\end{equation}
where
\begin{eqnarray}
\gamma_1=(acg-aef)/\alpha,\\
\gamma_2=(\beta de-cg+ef-ag-bdg)/\alpha^{1/2}.
\end{eqnarray}

It can be easily checked  that these solutions approach Freed's 
solutions in the limit $p_{nc}\rightarrow 0,\  p_c\rightarrow 0$.
In order that the steady state solutions for $P_+(n)$, $P_-(n)$ and 
$\Pi(n)$ are decreasing, rather than increasing, functions of $n$, 
we demand that $\alpha <1 $ which imposes the following constraint on 
the magnitudes of the parameters:
\begin{equation}
p_{-+}(p_c+\rho p_g)^2 < \rho p_{g}p_{d}p_{+-}
\label{eq-hybstab}
\end{equation} 
The condition (\ref{eq-hybstab}) reduces to the corresponding condition 
(\ref{eq-freedstab}) of the Freed model in the limit $p_c \rightarrow 0$.

\begin{figure}[h]
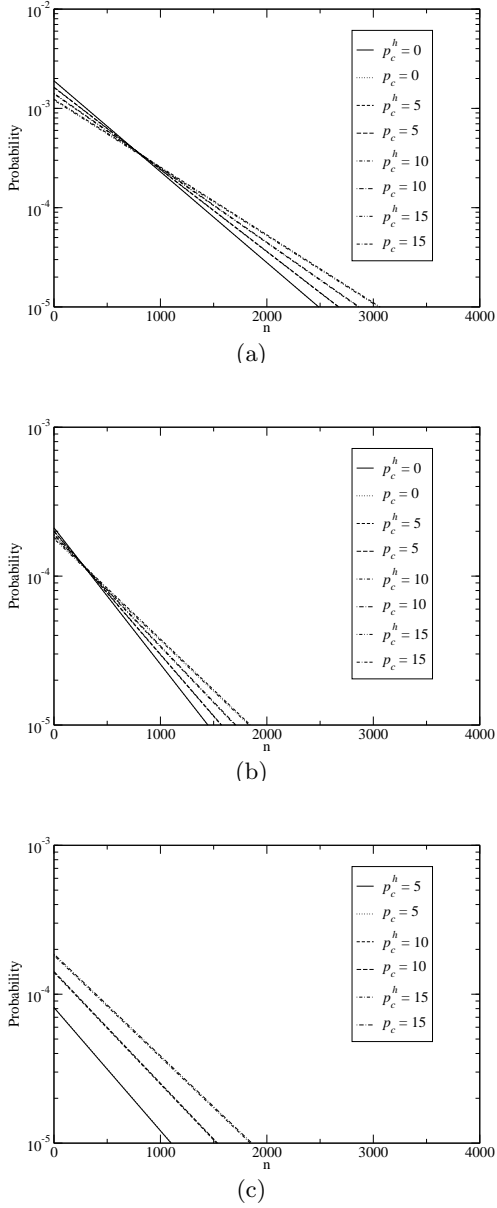

\begin{center}
\includegraphics[width=0.75\columnwidth]{fig3a.eps}
\vspace{.65cm}
\centerline{(a)}
\includegraphics[width=0.75\columnwidth]{fig3b.eps}
\vspace{.65cm}
\centerline{(b)}
\includegraphics[width=0.75\columnwidth]{fig3c.eps}
\vspace{.65cm}
\centerline{(c)}
\end{center}
\caption{The steady-state length distributions (a) $P_{+}(n)$,
(b) $P_{-}(n)$ and (c) $\Pi(n)$ of microtubules 
in our hybrid model for several values of $p_c$ and compared with 
corresponding distributions in our Hill-like model (with appropriate 
rescaling of the parameters and variables following (\ref{eq-rescale}); 
the common parameters are $p_g = p_n = 125$, $p_d=900$, $p_{-+}=0.08$, 
$p_{+-}=0.22$, $p_{nc}=p_c$, and $\rho=0.8$, $N_0=0.2$ (in respective 
units). }
\label{fig-hybrid}
\end{figure}

In order to plot the steady-state distributions (\ref{eq-hybppn}), 
(\ref{eq-hybpmn}) and (\ref{eq-hybpin}) in our hybrid Hill-Freed-like 
model and to compare these with the distributions 
(\ref{eq-hsol1}), (\ref{eq-hsol2}) and (\ref{eq-hsol3}) for the same 
set of parameters, we first converted the concentrations of the different 
types of microtubules into probabilities and, then, chose the numerical 
values of the parameters so as to satisfy the following relations:
\begin{eqnarray}
p^{h}_g = p_g \rho, \nonumber \\
p^{h}_d = p_d, \nonumber \\
p^{h}_c = p_c, \nonumber\\
p^{h}_{+-} = p_{+-}, \nonumber \\
p^{h}_{-+} = p_{-+}. 
\label{eq-rescale}
\end{eqnarray}
The distributions plotted in fig.\ref{fig-hybrid} shows that with 
proper rescaling of the parameters, as mentioned in (\ref{eq-rescale}), 
the results for the Hill-like model and the hybrid Hill-Freed-like 
model are almost identical. Moreover, the higher is the value of $p_c$, 
the longer are the tails of the distributions; this, as explained 
already in the context of the Hill-like model, is a consequence of the 
stronger suppression of the catastrophes by higher $p_c$.

The mean lengths of the microtubules, which correspond to the distributions 
plotted in the fig.\ref{fig-hybrid}, are plotted against $p_c$ in 
fig.\ref{fig-meanlength}. Surprisingly, for this set of parameter values 
the mean lengths of the microtubules tipped with drug-bound tubulin 
and those tipped with drug-free tubulins (both in the growing and 
shrinking phases) are identical for almost all values of $p_c$. However, 
even for this set of parameters values, the fraction of microtubules 
with drug-bound cap increases with increasing $p_c$ while that with 
drug-free cap decreases (see fig.\ref{fig-psum}).

\begin{figure}[h]
\begin{center}
\includegraphics[width=0.75\columnwidth]{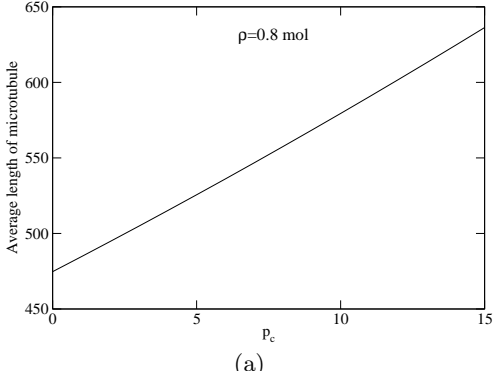}
\vspace{.65cm}
\centerline{(a)}
\includegraphics[width=0.75\columnwidth]{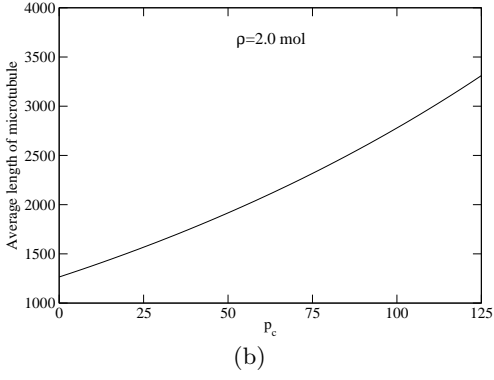}
\vspace{.65cm}
\centerline{(b)}
\end{center}
\caption{Mean length of the microtubules in the hybrid model. The numerical 
values of all the parameters, which are not shown explicitly, are identical 
to those in fig.\ref{fig-hybrid}. 
}
\label{fig-meanlength}
\end{figure}

\begin{figure}[h]
\begin{center}
\includegraphics[width=0.75\columnwidth]{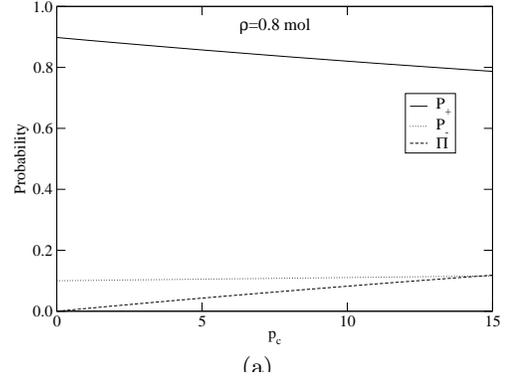}
\vspace{.65cm}
\centerline{(a)}
\includegraphics[width=0.75\columnwidth]{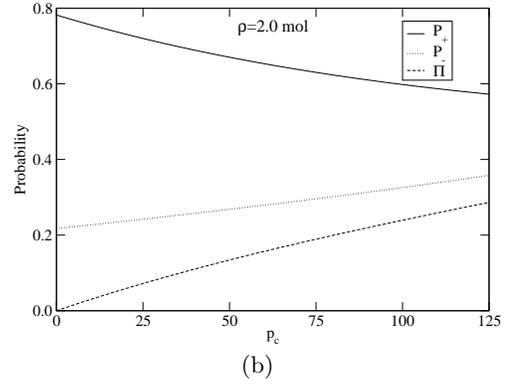}
\vspace{.65cm}
\centerline{(b)}
\end{center}
\caption{Fractions of the microtubules that are tipped with drug-free 
tubulin and are in the growing phase ($P_+$) or in the shrinking phase 
($P_-$) and those of microtubules tipped with durg-bound tubulin ($\Pi$).
The numerical values of all the parameters, which are not shown explicitly, 
are identical to those in the fig.\ref{fig-hybrid}.
}
\label{fig-psum}
\end{figure}

We now define the ``{\it effective}'' catastrophe frequency and the 
``{\it effective}'' rescue frequency by the relations 
\begin{equation}
p_{+-}^{eff} = \frac{p_{+-} P_+}{P_+ + P_- + \Pi} = \frac{p_{+-} P_+}{1-P_0}
\label{eq-effppm}
\end{equation}     
and                                                  
\begin{equation}
p_{-+}^{eff} = \frac{p_{-+} P_-}{P_+ + P_- + \Pi} = \frac{p_{-+} P_-}{1-P_0}
\label{eq-effpmp}
\end{equation}   
These ``effective'' frequencies of catastrophe and rescue are plotted 
against $p_c$ in fig.\ref{fig-efffreq}. This trend of variation is 
qualitatively similar to the corresponding results of laboratory 
experiments performed with the catastrophe-supprssing drug {\it vinblastine} 
\cite{panda1}.

\begin{figure}[h]
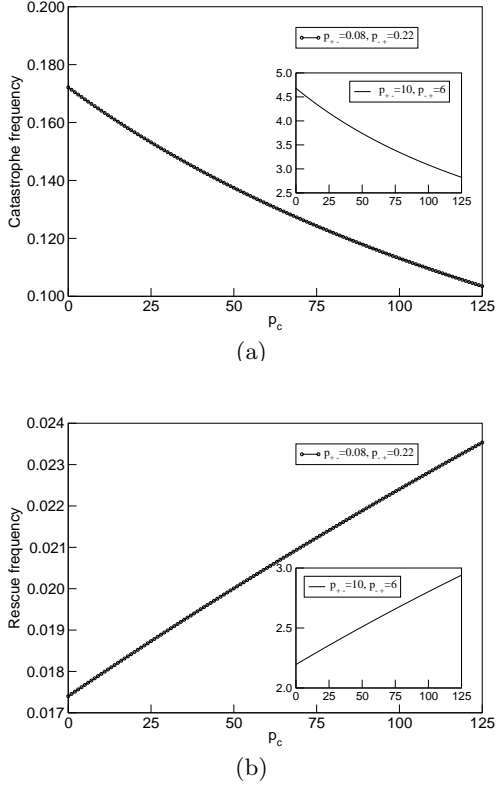

\begin{center}
\includegraphics[width=0.75\columnwidth]{fig6a.eps}
\vspace{.65cm}
\centerline{(a)}
\includegraphics[width=0.75\columnwidth]{fig6b.eps}
\vspace{.65cm}
\centerline{(b)}
\end{center}
\caption{Effective catastrophe frequency (a) and effective rescue 
frequency (b), defined through equations (\ref{eq-effppm}) and 
(\ref{eq-effpmp}), respectively, are plotted against $p_c$ for $\rho = 2.0$. 
Note the different parameters used for the insets.
The numerical values of all the parameters, which are not shown explicitly,
are identical to those in the fig.\ref{fig-hybrid}. 
}
\label{fig-efffreq}
\end{figure}

\subsection{Stability of the steady-state} 

We have obtained the exact analytical expressions for the matrices 
that decide the stability of the steady-state solutions, which we 
derived earlier in this section, against small deviations. For this 
purpose we expand the nonlinear kinetic equations about the steady  
states and retain only upto the terms linear in the deviations and 
drop all the terms containing higher orders of deviations.

Let us define the small deviations 
\begin{equation}
\delta \rho = \rho(t) - [\rho]_{ss}, 
\end{equation}
\begin{equation}
\delta P_{\pm}(n) = P_{\pm}(n,t) - [P_{\pm}]_{ss}  
\end{equation}
\begin{equation}
\delta \Pi (n) = \Pi (n,t) - [\Pi]_{ss}  
\end{equation}
from the corresponding steady-states where the steady-state values 
$[P_{\pm}]_{ss}$, $[\Pi]_{ss}$, etc. are given by the equations 
(\ref{eq-hybppn}), (\ref{eq-hybpp1}), (\ref{eq-hybpmn}), 
(\ref{eq-hybpm1}), (\ref{eq-hybpin}), (\ref{eq-hybpi1}), etc. 
We shall also use the symbols $[{\cal P}_+]_{ss}$, $[{\cal P}_-]_{ss}$ and 
$[{\Phi}]_{ss}$ to denote the steady-state values of $P_+$, $P_-$ and 
$\Pi$ given by the equations (\ref{eq-sumps}). 

Expanding the equations (\ref{eq-hyb1})- (\ref{eq-hyb6}) about the 
steady state and retaining upto the terms linear in the small 
deviations we get 
\begin{eqnarray} 
\frac{d \delta P_+(n)}{dt} = p_g \delta \rho \{[P_+(n-1)]_{ss} + [\Pi(n-1)]_{ss}\} \nonumber \\
+ p_g [\rho]_{ss} \{\delta P_+(n-1) + \delta \Pi(n-1)\} + p_{-+} \delta P_-(n) \nonumber \\
- p_g \delta \rho [P_+(n)]_{ss} - p_g [\rho]_{ss} \delta P_+(n) \nonumber \\ 
- (p_c + p_{+-}) \delta P_+(n)~~ ~~{\rm for}~~ n \geq 2
\end{eqnarray}
\begin{eqnarray} 
\frac{d \delta P_-(n)}{dt} = p_d \delta P_-(n+1) + p_{+-} \delta P_+(n) \nonumber \\
- (p_d + p_{-+}) \delta P_-(n) \nonumber \\
{\rm for}~~ n \geq 1
\end{eqnarray}
\begin{eqnarray} 
\frac{d \delta \Pi(n)}{dt} = p_c \{\delta \Pi (n-1) + \delta P_+(n-1)\} \nonumber \\
- p_g [\rho]_{ss} \delta \Pi(n) 
- p_g \delta \rho [\Pi(n)]_{ss}  - p_c \delta \Pi(n) \nonumber \\
{\rm for}~~ n \geq 2
\end{eqnarray}
\begin{eqnarray} 
\frac{d \delta P_+(1)}{dt} = p_n \tau + p_{-+} \delta P_-(1) \nonumber \\ 
- p_g \{[\rho]_{ss} \delta P_+(1) + \delta \rho [P_+(1)]_{ss}\} \nonumber \\ 
- (p_c + p_{+-}) \delta P_+(1) 
\end{eqnarray}
\begin{eqnarray} 
\frac{d \delta \Pi(1)}{dt} = p_{nc} \{- \sum_{n=1}^{\infty} \delta P_+(n) - \sum_{n=1}^{\infty} \delta P_-(n) - \sum_{n=1}^{\infty} \delta \Pi(n)\} \nonumber \\ 
- p_g \{[\rho]_{ss} \delta \Pi(1) + \delta \rho [\Pi(1)]_{ss} \} - p_c \delta \Pi(1) \nonumber \\
\end{eqnarray}
\begin{eqnarray} 
\frac{d \delta \rho}{dt} = - p_n \tau - p_g \delta \rho \{\sum_{n=1}^{\infty} [P_+(n)]_{ss} + \sum_{n=1}^{\infty} [\Pi(n)]_{ss}\} \nonumber \\ 
- p_g [\rho]_{ss} \{\sum_{n=1}^{\infty} \delta P_+(n) + \sum_{n=1}^{\infty} \delta \Pi(n)\} \nonumber \\
+ p_d \sum_{n=1}^{\infty} \delta P_-(n)
\end{eqnarray}
where 
\begin{eqnarray} 
\tau = \delta \rho \{N_0 - [{\cal P}_+]_{ss} - [{\cal P}_-]_{ss} - [\Phi]_{ss}\} \nonumber \\ 
+ [\rho]_{ss} \{ - \sum_{n=1}^{\infty} \delta P_+(n) - \sum_{n=1}^{\infty} \delta P_-(n) - \sum_{n=1}^{\infty} \delta \Pi(n) \} \nonumber \\
\end{eqnarray}

Next, we define 
\begin{eqnarray} 
\Delta_{\pm} = \sum_{n=1}^{\infty} \delta P_{\pm}(n) \\
\Delta_0 = \sum_{n=1}^{\infty} \delta \Pi(n) 
\end{eqnarray}
and the column vectors 
\begin{equation}
{\bf V}(t) = \left( \begin{array}{lr} \Delta_+\\
                           \Delta_- \\
                           \Delta_0 \\
                           \delta \rho 
             \end{array} \right)
\end{equation}
\begin{equation}
{\bf N} = \left( \begin{array}{lr} 0\\
                           - p_d \\
                               0 \\
                               0 
          \end{array} \right)
\end{equation}
The equations obtained from the linear stability analysis above 
can be written as 
\begin{equation}
\frac{d{\bf V}(t)}{dt} = {\bf M} {\bf V}(t) + {\bf N} \delta P_-(1) 
\end{equation}
where the matrix ${\bf M}$ is given by 
\begin{widetext}
\begin{equation}
{\bf M} = \left( \begin{array}{llcl} - (p_c + p_{+-})-p_n[\rho]_{ss}~~ & p_{-+} - p_n[\rho]_{ss}~~ & (p_g-p_n)[\rho]_{ss}~~ & {\bf M}_{14}\\
                           p_{+-}~~ & - p_{-+}~~ & 0~~ & 0 \\
                           p_c-p_{nc}~~ & - p_{nc}~~ & -p_g[\rho]_{ss}-p_{nc}~~ & -p_g [\Phi]_{ss} \\
                           (p_n-p_g)[\rho]_{ss}~~ & p_n[\rho]_{ss}+p_d~~ & (p_n-p_g)[\rho]_{ss}~~ & {\bf M}_{44} 
          \end{array} \right)
\end{equation}
\end{widetext}
where 
\begin{equation}
{\bf M}_{14} = p_g [\Phi]_{ss}+p_n\{N_0-[{\cal P}_+]_{ss}-[{\cal P}_-]_{ss}-[\Phi]_{ss}\}
\end{equation}
\begin{equation}
{\bf M}_{44} = (p_n-p_g)([{\cal P}_+]_{ss}+[\Phi]_{ss})-p_n(N_0-[{\cal P}_-]_{ss})
\end{equation}

The characteristic equation is quartic in the eigenvalue $\lambda$. 
We have systematically investigated the trend of variation of the 
eigenvalues ($m_1, m_2, m_3, m_4$) of the matrix ${\bf M}$ with the 
variation of $p_c$ for several sets of values of the other parameters; 
those for two values of $\rho$ are given in table \ref{tab-2}. In both  
these cases, there is a range $0 \leq p_c \leq p_c^{max}$ of 
values of $p_c$ where all the eigenvalues remain negative indicating 
stability of the steady state. Moreover, the higher is the steady-state 
density $\rho$ of the free tubulins in the solution larger is the value 
of $p_c^{max}$. As $p_c$ is incresed beyond $p_c^{max}$, one of the 
negative eigenvalues simply changes sign which indicates instability of 
the steady-state. The physical reason for the instability of the steady 
state at sufficiently high $p_c$ is that the higher is the $p_c$ the 
larger is the fraction of microtubules tipped with drug-bound tubulin 
which can only grow but cannot shrink.

\vspace{1cm}

\begin{table}
\begin{tabular}{|c|c|c|c|c|} \hline
\multicolumn{5}{|c|}{$\rho=0.8M$} \\ \hline
 $p_c$   &$m_1$    &  $m_2$   &   $m_3$  &  $m_4$ \\ \hline
   0     &-99.998  & -0.295   & -22.506  & -100.000 \\ \hline
   5     &-105.000 & -0.185   & -22.633  & -104.982 \\ \hline
   10    &-110.000 & -0.094   & -22.734   & -109.972 \\ \hline
   15    &-115.000 & -0.019   & -22.814   & -114.967 \\ \hline
   16    &-116.000 & -0.005   & -22.828   & -115.967 \\ \hline
   17     &-117.000 & 0.008    & -22.842   & -116.966 \\ \hline
\multicolumn{5}{|c|}{$\rho=2.0M$} \\ \hline
$p_c$  & $m_1$ &   $m_2$  &   $m_3$  &  $m_4$ \\ \hline
   0   &-249.999 & -0.299 & -19.566 & -250.000 \\ \hline
  50   &-300.000 & -0.121 & -19.717 & -300.023 \\ \hline
  100   &-350.000 & -0.030 & -19.785 & -350.051 \\ \hline
  126   &-376.000 & -0.001 & -19.803 & -376.061 \\ \hline
  127   &-377.000 &  0.002 & -19.804 & -377.062 \\ \hline
\end{tabular}
\\
\caption{The eigenvalues of the linear stability matrix ${\bf M}$. 
The common parameters are 
$p_g=p_n=125, p_d=900, p_{-+}=0.08, p_{+-}=0.22, p_{nc}=p_{c}, N_0=0.2$ 
(in respective units).  } 
\label{tab-2}
\end{table}

\vspace{1cm}

Thus, although the steady-state distributions of the microtubules in 
the Hill-like model and those in the hybrid Hill-Freed-like model 
can be made practically identical by making appropriate correspondence 
of the parameters in the two models, the latter has a richer dynamics 
which is reflected also in the linear stability analysis.

\section{\label{sec5} Summary and conclusion}

In this paper we have developed models for studying some generic 
effects of a class of drugs on the polymerization-depolymerization 
dynamics of microtubules in the absence of GTP and GDP. The class 
of generic drugs under consideration suppress catastrophes; more 
specifically, we assumed that (i) the microtubules capped with 
drug-bound tubulins do not exhibit catastrophe, and (ii) the rate 
constant for the attachment of a drug-bound tubulin is, in general, 
different from that of a drug-free tubulin. Although the effects of 
real catastrophe-suppressing drugs are known \cite{jordan} to be 
more complicated than those of the generic model assumed here, some 
predictions of the model are in good qualitative agreement with the 
corresponding effects of {\it vinblastine} over a limited parameter 
regime.  

One of the two models, namely, the Hill-like model proposed here is 
an extension of the model developed by Hill \cite{hill} 
to describe microtubule polymerization-depolymerization dynamics in 
the absence of drugs. Although mathematical treatment of this model 
is quite simple, it does not take explicitly account for the dynamics 
of the tubulin concentration in the solution. Therefore, we have 
also developed a more detailed model in which the effects of the 
concentration of the pure (i.e., drug-free) tubulin subunits are taken 
into account in our model in a manner similar to that done by Freed 
\cite{freed} in his recent theoretical study of the microtubule dynamics 
in the absence of drugs. However, in this model the effects of the 
tubulin subunits bound to the drug molecules could not be taken 
into account in a similar manner. 

The reason for the difficulty in treating the concentrations of the 
drug-free and drug-bound tubulins in solution on an equal footing 
is generic to the Hill-Freed approach. In order that the dynamical 
equation for the concentration of drug-bound tubulin subunits can reach 
a steady-state, we have to allow both attachment as well as detachment 
of the drug-bound subunits to the microtubule. However, if shrinking 
of a microtubule tipped with drug-bound tubulin is allowed, then, 
immediately after such a detachment, one needs to know the status of 
the subunit at the newly exposed tip, i.e., whether the tip consists of 
a drug-free tubulin or a drug-bound tubulin. But, in the Hill-Freed 
approach, the system does not have a memory of the past history in the 
sense that the the model does not keep a record of the status (i.e., 
whether or not bound to a drug molecule) of {\it all} the tubulin 
subunits, starting from the nucleation center up to the tip. 

Therefore, in order to overcome this technical difficulty, we have 
incorporated the effects of the drug-bound tubulin subunits in a 
manner similar to the approach followed in the Hill model \cite{hill}. 
Thus, our second model may be regarded as a hybrid of the Hill-like and 
Freed-like modeling strategies for the concentrations of the tubulin 
subunits.

For both the Hill-like model and the Hill-Freed-like hybrid model 
we have derived exact analytical expressions for the steady-state 
probability distributions of the lengths of microtubules tipped with 
drug-bound tubulin subunits as well as those of microtubules tipped 
with pure (i.e., drug-free) tubulin subunits in the growing and 
shrinking phases. We have also compared the trends of variations 
of some of the relevant quantities with the variation of the dosage 
of the catastrophe-suppressing drug.

We have carried out linear stability analysis of the steady-states and 
established that in both the models the length distributions of the 
microtubules remain stable unless $p_c$ becomes sufficiently high to 
destabilize the steady-state.  

\vspace{0.5cm}

\appendix
\section{On the use of Chebyshev polynomial}
\label{appA}
It is straight forward to see that 
\begin{eqnarray}
a' + c' + b'd' &=& 1 + \frac{p^{F}_g \rho (p^{F}_d + p^{F}_{-+})}{p^{F}_d (p^{F}_g \rho + p^{F}_{+-})} \nonumber \\
&=& 1 + a' c'
\end{eqnarray}
Therefore, re-expressing $\lambda'$ as 
\begin{equation}
\lambda' = \frac{1 + a'c'}{2(a'c')^{1/2}} 
= \frac{1}{2}\biggl[\biggl\{(a'c')^{1/4} - (a'c')^{-1/4}\biggr\}^2 + 2 \biggr]
\end{equation}
we find, in general, $\lambda' > 1$.

\section{steady-state solutions of  Hill-like model with drugs}
\label{appB}
Here we obtain the steady-state solutions of the equations 
(\ref{eq-htype1})-(\ref{eq-htype6}) for Hill-like model in the 
presence of drugs by equating the RHS of these equations to zero. 
The solution for  $P_-(1)$ is obtained by demanding $dP(0,t)/dt=0$. 
This leads to 
\begin{eqnarray}
P_-(1)=\frac{(p_g+p_c)}{p_d} P(0)=y P(0) 
\end{eqnarray}
where $y$ is given by the equation (\ref{eq-htypey}). 
Similarly, claiming $dP_+(1,t)/dt=0$ and $d\Pi(1,t)/dt=0$, we have 
\begin{eqnarray}
&&P_+(1)= \frac{[p_g p_d+p_{-+}(p_g+p_c)]}{p_d(p_g+p_c+p_{+-})} P(0) = x P(0)\\
&& \Pi(1)=\frac{p_c}{p_c+p_g} P(0)=z P(0).
\end{eqnarray} 
where $x$ and $z$ are given by the equations (\ref{eq-htypex}) and 
(\ref{eq-htypez}), respectively.
The special case corresponding to $n=1$ of the equation $dP_(n,t)/dt = 0$, 
leads to the steady state solution for $P_-(2)$,
\begin{eqnarray}
P_-(2)=\frac{1}{p_d}[(p_d+p_{-+}) y-p_{+-} x]P(0).\\ 
\end{eqnarray}
A little bit of algebraic manipulqation leads to 
\begin{eqnarray}
P_-(2)=y(x+z) P(0).
\end{eqnarray}
Steady-state solutions for  $P_+(2)$ and $\Pi(2)$ can be obtained
in a similar way. Steady-state solutions for $P_-(n)$, $P_+(n)$ 
and $\Pi(n)$, for arbitrary $n$, are straightforward generalizations 
of our observations upto $n=4$. 

Since $P_+(n)$, $P_-(n)$ and $\Pi(n)$ are probabilities of finding 
microtubules of $n$ subunits in growing, shrinking and in 
catastrophe-arrested phase, we expect the following normalization
condition
\begin{eqnarray}
P(0)+\sum_{n=1}^{\infty}(P_+(n)+P_-(n)+\Pi(n))=1.
\end{eqnarray}
This leads to the form (\ref{eq-hsol4}) for $P(0)$. 


\begin{acknowledgments}
We thank Balaji Prakash for fruitful discussions and Karl Freed for 
useful comments on a preliminary draft of the manuscript. 
\end{acknowledgments}



\begin{thebibliography}{99}
\bibitem{chowetal} There are some errors in the formulae in the 
ref.\cite{freed}; here we give the corresponding corrected expressions.\\[0.2cm]
\end{thebibliography}
\end{document}